\documentclass[twocolumn,showpacs,%
  nofootinbib,aps,superscriptaddress,%
  prd,notitlepage,showkeys,10pt]{revtex4-1}

\usepackage{amssymb}
\usepackage{amsmath}
\usepackage{physics}
\usepackage{graphicx}
\usepackage{dcolumn}
\usepackage{hyperref}
\usepackage{xcolor}
\usepackage[normalem]{ulem}
\usepackage{siunitx}

\DeclareMathAlphabet\mathbfcal{OMS}{cmsy}{b}{n}

\newcommand{\pC}{p_{\text{C}}}
\newcommand{\mpC}{\bar{p}_{\text{C}}}
\newcommand{\Fsa}{\mathbf{F}_{\text{sa}}}
\newcommand{\Fad}{\mathbf{F}_{\text{ad}}}
\newcommand{\Fzad}{\mathbf{F}_{\text{z-ad}}}
\newcommand{\thetam}{\text{acos}\left( 1/\sqrt{3}\right)}

\begin{document}

\title{Mean-field Pulse Adaptation for the Circularization of Interacting Rydberg Atoms}

\author{Matthias Hüls}
\affiliation{Forschungszentrum Jülich GmbH, Peter Grünberg Institute, Quantum Control (PGI-8), 52425 Jülich, Germany}
\affiliation{Institute for Theoretical Physics, University of Cologne, 50937 Köln, Germany}

\author{Felix Motzoi}
\affiliation{Forschungszentrum Jülich GmbH, Peter Grünberg Institute, Quantum Control (PGI-8), 52425 Jülich, Germany}
\affiliation{Institute for Theoretical Physics, University of Cologne, 50937 Köln, Germany}

\author{Tommaso Calarco}
\affiliation{Forschungszentrum Jülich GmbH, Peter Grünberg Institute, Quantum Control (PGI-8), 52425 Jülich, Germany}
\affiliation{Institute for Theoretical Physics, University of Cologne, 50937 Köln, Germany}
\affiliation{Dipartimento di Fisica e Astronomia, Università di Bologna, 40127 Bologna, Italy}

\author{Eloisa Cuestas}
\affiliation{Forschungszentrum Jülich GmbH, Peter Grünberg Institute, Quantum Control (PGI-8), 52425 Jülich, Germany}
\affiliation{OIST Graduate University, Onna, Okinawa, Japan}

\begin{abstract}
Arrays of circular Rydberg atoms provide a promising platform for quantum simulation and computation; however, their preparation in the presence of interatomic interactions remains a major challenge. While optimal control methods have enabled the design of fast and accurate radio-frequency pulses for the circularization of a single atom and of an atom pair, the extension to more atoms is fundamentally limited by the exponential growth of the Hilbert space, which renders numerical simulations computationally infeasible. Here, we introduce an effective model that treats interactions within a mean-field approximation, thereby enabling the simulation of large atomic systems. Our model further enables the adaptation of pulses optimized for non-interacting atoms to interacting systems, based on the computation of a single time evolution. For two interacting $^{87}$Rb atoms, we demonstrate that the error of our method remains below $\SI{1}{\percent}$ and that our adapted pulses recover the initial performance of optimal pulses in the regime of weak to moderate interaction strengths.
\end{abstract}

\maketitle

Atoms in circular Rydberg states, where one valence electron is excited to the maximal orbital angular momentum state for a given large principal quantum number $n$, feature long lifetimes and strong interatomic interactions \cite{Hlzl2024, Wu2023, Ravon2023}, rendering them a promising platform for quantum computing \cite{Cohen2021}, simulation \cite{Nguyen2018, Meinert2020} and sensing \cite{Facon2016, Dietsche2019}. The final part of their preparation, known as \textit{circularization}, involves the transfer from a state with low magnetic quantum number $m$ to the circular state with maximal $m=n-1$ by coupling the atom to a circularly polarized radio-frequency (RF) field. This step can be implemented via an adiabatic passage, in which a slowly ramped electric field tunes the system through the resonance with the RF field \cite{Hlzl2024, Ravon2023, Nussenzveig1993}. A faster circularization is achieved by short pulses designed with pulse shaping algorithms \cite{Signoles2017, Patsch2018}. For non-interacting $^{87}$Rb atoms in $n=52$ state manifold, pulses with durations of around \SI{100}{\nano\second}, designed to achieve a circularization probability of \SI{99}{\percent} in a simulation, yielded a probability of \SI{96.2}{\percent} in an experimental implementation \cite{Larrouy2020}. The corresponding experiment was performed with a dilute atomic cloud, where interatomic interactions can be neglected.

Applications for quantum simulation and computation rely on arrays of interacting circular Rydberg atoms \cite{Cohen2021, Nguyen2018, Meinert2020}. However, interatomic interactions severely limit the efficiency of adiabatic methods through relaxation effects, confining their applicability to atomic arrays with negligible interactions \cite{mehai2023, Mhaignerie2025}. This limitation can be overcome with fast pulses. For a pair of atoms, pulses optimized for non-interacting atoms can be adapted to the interactions through a phase modulation that compensates for interaction-induced frequency shifts \cite{Huels26}. Adapted pulses prepare two $^{87}$Rb atoms in a simultaneous $n=52$ circular state with a probability above $\SI{95}{\percent}$ for all angular configurations and interatomic distances down to $\SI{6.5}{\micro\meter}$. In this work, we generalize the pulse adaptation approach by extending it to arbitrary arrays of $N$ atoms and by including amplitude corrections, therefore increasing its performance. The proposed adaptation scheme is based on an effective semi-classical model that resolves the $d$-dimensional state space of each atom while treating the interactions as classical fields. The effective model can be simulated in a state space of dimension $Nd$, a drastic reduction from the $d^N$ required for an exact product-basis simulation. In the regime of weak to moderate interactions, where entanglement effects between the atoms can be neglected, its imprecision with respect to the exact simulation remains below $\SI{1}{\percent}$.

Our aim is to excite the valence electron of alkali atoms to a circular state $\ket{n\text{C}}$. 
Following Refs. \cite{Patsch2018, Huels26} we consider atoms subject to static electric and magnetic fields $\mathbf{B} = B \mathbf{e}_z$ and $\mathbfcal{E} = \mathcal{E} \mathbf{e}_z$, that lift degeneracies and define the quantization axis. During the circularization, the atom is coupled to a radio-frequency (RF) pulse $\mathbf{F}(t)$ through $\hat{h}_{\text{RF}}(t) = -\hat{\mathbf{d}} \cdot \mathbf{F}(t)$, yielding the total Hamiltonian
\begin{align}
\label{eq:single atom hamiltonian}
    \hat{h}(t) = \hat{h}_s + \hat{h}_{\text{RF}}(t),
\end{align}
where $\hat{h}_s = \hat{h}_0 - \hat{{\mathbf{d}}} \cdot \mathbfcal{E} + \mu_{B}\mathbf{B} \cdot \hat{\mathbf{L}}$ is the static part of the Hamiltonian. We neglect spin-orbit coupling. $-e$ is the electron charge, $\mu_{B}$ the Bohr magneton, $\hat{h}_0$ the atomic Hamiltonian, $\hat{\mathbf{d}}$ the dipole operator, and $\hat{\mathbf{L}}$ the orbital angular momentum operator.

The pulse $\mathbf{F}(t)$ is $\sigma_+$ polarized, i.e. its components $F^x(t)$ and $F^y(t)$ oscillate with a phase difference of $\pi/2$. In the ideal case, it transfers the entire population from an initial low-$m$ state to the circular state with maximal $m=n-1$. However, because the pulse must fulfill experimental constraints and account for anharmonicities --- primarily arising from quantum defects \cite{gallagher_1994_book} --- the design of pulses optimal for a circularization is nontrivial \cite{Patsch2018, Larrouy2020}. We take a pulse $\Fsa (t)$ that meets these requirements as given (see Refs. \cite{Patsch2018, Huels26} for details). $\Fsa (t)$ prepares a single atom in a final state $\ket{\psi(t_f)}$ close to a circular state, e.g., yields a circular state probability of $\pC = \abs{\bra{\psi(t_f)}\ket{n\text{C}}} = \SI{99}{\percent}$. Our goal is to recover this performance for interacting atoms by adapting $\Fsa (t)$ accordingly.

We generalize the settings to $N$ interacting atoms described by
\begin{align}
\label{Eq: N Atom general Hamiltonian}
\hat{H} = \sum_{i=1}^N \left[ \hat{H}_{s, i} + \hat{H}_{\text{RF}, i}(t)\right] + \hat{H}_{\text{int}},
\end{align}
where we assume the static fields $\mathbfcal{E}$ and $\mathbf{B}$ and pulse $\mathbf{F}(t)$ to be uniform over the extent of the atom arrangement and to couple to each atom locally without delays, i.e. $\hat{H}_{s,i} = \hat{\mathbb{I}}^{\otimes (i-1)} \otimes \hat{h}_s \otimes \hat{\mathbb{I}}^{\otimes (N-i)}$ and similar for $\hat{H}_{\text{RF}, i}$. All atom pairs $\left\{(i,j)\right\}$ interact through dipole-dipole coupling
\begin{small}
\begin{align}
\label{Eq: N Atom exact interaction Hamiltonian}
\hat{H}_{\text{int}} = \frac{1}{4\pi \epsilon_0}\sum_{i}^N \sum_{j > i}^N \frac{1}{R_{ij}^3}\left[ \hat{\mathbf{d}}_i \cdot \hat{\mathbf{d}}_j - 3 \left(\hat{\mathbf{d}}_i \cdot \mathbf{e}_{ij} \right) \left( \hat{\mathbf{d}}_j \cdot \mathbf{e}_{ij}\right)\right],
\end{align}
\end{small}
 where $\epsilon_0$ is the vacuum permittivity and $\mathbf{R}_{ij} = R_{ij}\mathbf{e}_{ij}$ is the interatomic vector between atoms $i$ and $j$. An exact simulation of the multi-atom states $\ket{\Psi(t)}$ requires a state space dimension $D = d^N$, which, even though it can be reduced by exploiting symmetries or neglecting barely populated states \cite{Huels26}, remains significantly larger than the state space dimension $d$ required to simulate a single atom. Consequently, exact simulations and thereupon optimizations of the circularization for a growing number $N$ of atoms become computationally infeasible quickly. In order to still access these systems through a simulation, we propose an effective model.

Following a semi-classical approach, we simulate the $N$ atoms individually, resolving their respective state space of dimension $d$, while modeling their interactions by classical fields. This model is valid within a Hartree approximation, in which the multi-atom state $\ket{\Psi(t)}$ is approximated by a product state at any time $t$ during the evolution $\ket{\Psi(t)} \approx \bigotimes_{i=1}^{N} \ket{\psi_i(t)}$, where $\ket{\psi_i(t)}$ is the state of atom $i$. Therefore, the effective model renders accurate results only when the entanglement among atoms can be neglected. 

We focus on atom $i$ and model the interactions with other atoms $j \neq i$ in a mean-field approach. Electric dipole operators acting on atoms $j$ are replaced by expectation values, yielding the effective interaction Hamiltonian
\begin{align}
\label{Eq: N Atom effective interaction Hamiltonian}
\hat{h}_{\text{int}, i} =& - \hat{\mathbf{d}}_i \cdot \sum_{j \neq i} \frac{1}{4 \pi \epsilon_0 R_{ij}^3}\left[\langle \hat{\mathbf{d}}_j\rangle - 3 \left(\langle \hat{\mathbf{d}}_j\rangle  \cdot  \mathbf{e}_{ij} \right) \mathbf{e}_{ij}\right] \notag \\
=& - \hat{\mathbf{d}}_i \cdot \mathbf{I}_{i}(t).
\end{align}
In this picture, atom $i$ evolves in a classical interaction field $\mathbf{I}_{i}(t)$ that is generated by the mean dipole moments $\langle \hat{\mathbf{d}}_j\rangle = \bra{\psi_j(t)}\hat{\mathbf{d}}\ket{\psi_j(t)}$ of atoms $j$, which are understood as time-dependent. In the effective model, atom $i$ evolves according to 
\begin{align}
    \label{Eq: full effective Hamiltonian}
    \hat{h}_{\text{eff}, i} (t)= \hat{h}_s - \hat{\mathbf{d}} \cdot \left[\mathbf{F}(t) + \boldsymbol{I}_i(t)\right].
\end{align}
The time evolution of each atom is coupled to that of the others, requiring a simultaneous computation. Therefore, we discretize the evolution into time steps $t_k$ of duration $\Delta t$. At step $k$, we use the states $\ket{\psi_{j}(t_k)}$ in order to compute $\boldsymbol{I}_{i}(t_k)$ and correspondingly $\hat{h}_{\text{eff}, i}(t_k)$ for all $i$. We propagate the states according to the time-dependent Schrödinger equation
\begin{align}
    \label{eq: time propagation}
    \ket{\psi_{i}(t_{k+1})} = e^{-\frac{i}{\hbar} \Delta t \hat{h}_{\text{eff}, i}(t_k)} \ket{\psi_{i}(t_{k})}, 
\end{align}
where $\hbar$ is the reduced Planck constant. As we compute the evolutions of all atoms in parallel, the required state space dimension is reduced exponentially from $d^N$ (exact $N$-body simulation) to $Nd$.

\begin{figure*}[t]
 \centering
 \includegraphics[width=\textwidth]{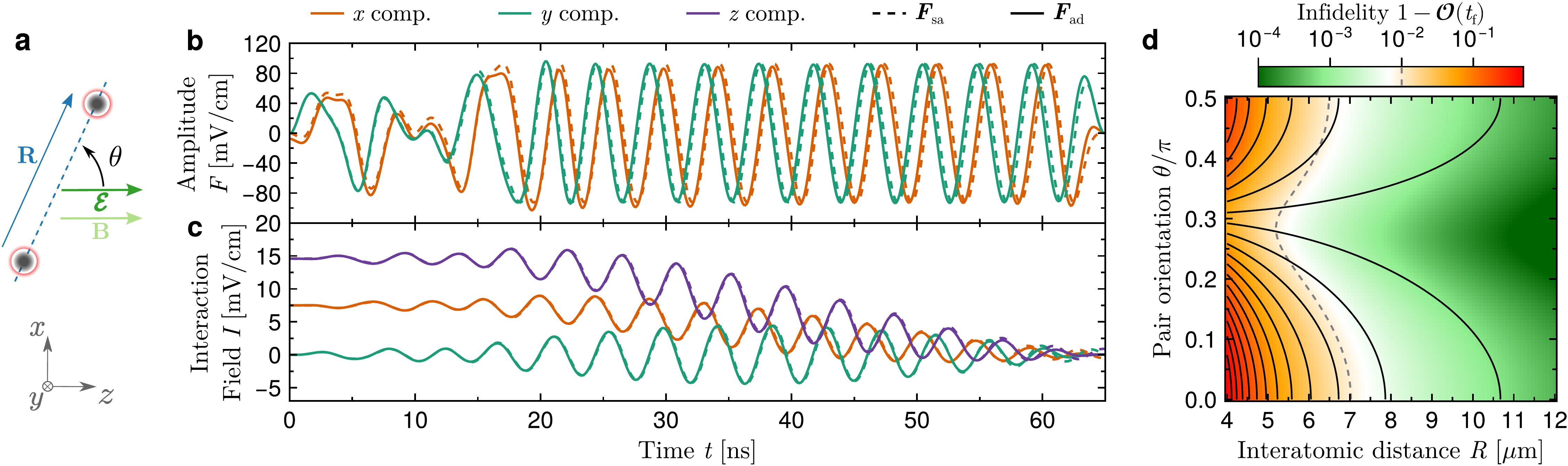}
   \caption{(a) Spatial arrangement of the atom pair. The two atoms are separated by the interatomic vector $\mathbf{R}$, which is oriented at an angle $\theta$ with respect to the static fields $\mathbfcal{E}$ and $\mathbf{B}$. (b) $x$- and $y$-components of the pulse $\Fsa(t)$ optimized for a single non-interacting atom (dashed curves). (c) A pair of atoms arranged at an interatomic distance $R = \SI{7}{\micro\meter}$ and angle $\theta = \SI{0.1}{\pi}$ and evolving under $\Fsa(t)$ can be efficiently simulated in the semi-classical effective model, where the interactions give rise to classical fields $\mathbf{I}(t)$ that couple to individual atoms. We use $\mathbf{I}(t)$ to add suitable amplitude and phase corrections to $\Fsa(t)$ therefore obtaining the adapted pulse $\Fad (t)$ --solid lines in panel (b). (d) The effective model is compared against exact two-body simulation via the overlap $\mathcal{O}$ defined in Eq.~\eqref{eq: fidelity}. For atom pairs at different arrangements $(R, \theta)$ the infidelity of the final state $1 - \mathcal{O}(t_{\text{f}})$ grows with the interaction strength dominated by the term $\propto \langle \hat{z}_j\rangle$ of Eq.~\eqref{eq:interaction_field_z} whose scaling is depicted as black contour lines. The gray-dashed contour curve indicates arrangements with an infidelity of $\SI{1}{\percent}$.}
 \label{fig: effective_model}
\end{figure*}

To illustrate our method and compare it with accurate numerical simulations, we revisit the circularization of two $^{87}$Rb atoms in $n=52$ Rydberg states \cite{Huels26}. The static fields are set to $\mathcal{E} = \SI{2.1}{\volt\per\centi\meter}$ and $B = 14 \, \text{G}$. The atoms are separated by $R = \SI{7}{\micro\meter}$ and oriented in the $x-z$ plane at an angle $\theta = \SI{0.1}{\pi}$ with respect to the $z$-axis, see Fig.~\ref{fig: effective_model} (a). For further details on the simulation, see \cite{Huels26}. We drive the pair with the pulse $\Fsa(t)$, which is optimized for a single atom and shown in Fig.~\ref{fig: effective_model} (b). Note that in the case of an atom pair coupled via dipole-dipole interactions, the system is symmetric under a permutation of the atoms \cite{Weber2017}. The atoms evolve identically $\ket{\psi_i(t)} = \ket{\psi_j(t)}$ and are prepared in circular states\footnote{In the effective model, the probability of preparing both atoms in a circular state factorizes as $p_{\text{CC}} = p_{\text{C},1} p_{\text{C},2}$. In an exact simulation, interatomic correlations typically prevent such a factorization.} with a mean probability $\mpC$ identical to the individual circular state probabilities, $\mpC = \frac{1}{N}\sum_i p_{\text{C},i} = p_{\text{C},1} = p_{\text{C},2}$. It thus suffices to simulate only one atom coupling to itself, with a corresponding state space of dimension $d$. The classical interaction fields $\mathbf{I}_1(t) = \mathbf{I}_2(t) = \mathbf{I}(t)$ are given by
\begin{small}
\begin{align}
    I^x(t) &= \frac{1}{4\pi\epsilon_0 R^3}\left( \langle\hat{d}^x\rangle (1 - 3 \sin^2\theta)  - 3 \sin \theta \cos \theta \langle\hat{d}^z \rangle\right) 
    \label{eq:interaction_field_x} \\
    I^y(t) &= \frac{1}{4\pi\epsilon_0 R^3} \langle\hat{d}^y\rangle
    \label{eq:interaction_field_y} \\
    I^z(t) &= \frac{1}{4\pi\epsilon_0 R^3}\left( \langle\hat{d}^z\rangle (1 - 3 \cos^2\theta)  - 3 \sin \theta \cos \theta \langle\hat{d}^x \rangle\right) . 
    \label{eq:interaction_field_z}
\end{align}
\end{small}
During its evolution, shown in Fig.~\ref{fig: effective_model} (c), $\mathbf{I}(t)$ has monotonically evolving and oscillatory contributions corresponding to the terms $\propto \langle \hat{d}^z \rangle $ and~$\propto \langle \hat{d}^x \rangle $~,~$\langle \hat{d}^y \rangle $ in Eqs.~\eqref{eq:interaction_field_x}--\eqref{eq:interaction_field_z} respectively.

We estimate the fidelity of the effective model by comparing the generated time evolution $\ket{\psi_{\text{eff}, i}(t)}$ with an exact two-body evolution $\ket{\Psi(t)}$ through the overlap 
\begin{align}
    \label{eq: fidelity}
    \mathcal{O}_i(t) = \ev{\rho_{i}(t)}{\psi_{\text{eff}, i}(t)},
\end{align}
where $\rho_{i}(t)$ is obtained by tracing out atom $j$ from $\ket{\Psi(t)}$. Due to the permutation symmetry, we have $\mathcal{O}_1 = \mathcal{O}_2 = \mathcal{O}$. At the end of the circularization, the infidelity reaches $1 - \mathcal{O}(t_{f}) = \SI{0.8}{\percent}$. As expected, the infidelity increases with interaction strength as can be seen in Fig.~\ref{fig: effective_model} (d) where we compare final infidelities obtained for different atom pair arrangements $(R, \, \theta)$ propagated with $\Fsa(t)$. For moderate interactions in the regime $R \geq \SI{7}{\micro\meter}$, the infidelity remains below $\SI{1.0}{\percent}$ for all angles. For stronger interactions the infidelity quickly grows, reaching $\SI{46.7}{\percent}$ for the arrangement $R = \SI{4}{\micro\meter}$ and $\theta = 0$. 

Overall, the strength of the dipole-dipole interactions is governed by the coupling of the $z$-components of the atomic dipole operators. This dominance is evidenced by the infidelity following the scaling $(1-3\cos^2\theta)/R^3$ \cite{Ravets2015}, shown as contour lines in Fig.~\ref{fig: effective_model} (d). Note that, beyond the static arrangements considered here, the effective model could also enable efficient simulations of the circularization for moving atoms, thereby allowing the generation of flying qubits for fast array formation and defect correction in dynamical architectures \cite{Hwang23}. 

\begin{figure*}[t]
\centering
\includegraphics[width=\textwidth]{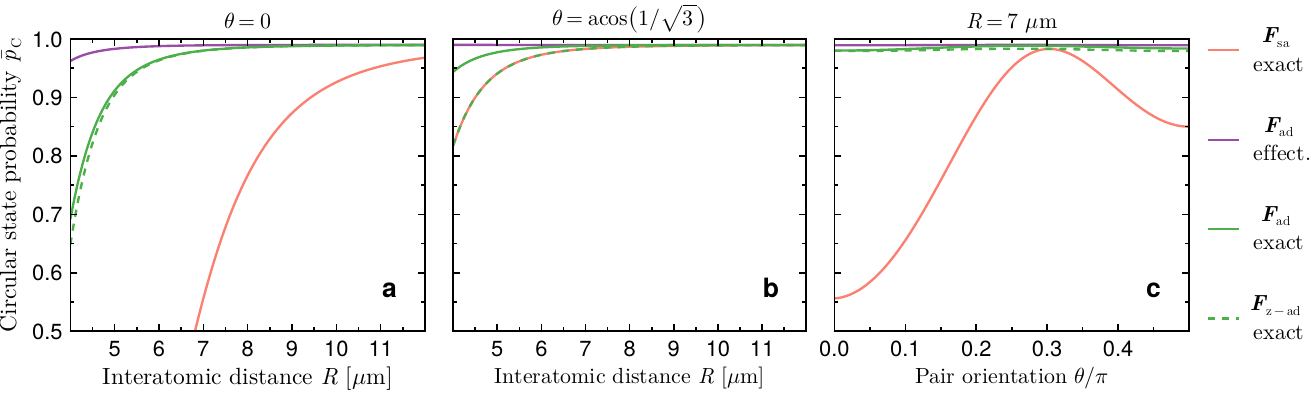}
   \caption{Mean circular state probability $\mpC$ obtained via an exact or effective simulation of an atom pair for different interatomic distances $R$ and fixed orientations $\theta = 0$ (a) or $\theta = \text{acos}\left(1/\sqrt{3} \right)$ (b) with respect to the static electric and magnetic fields, or different $\theta$ and fixed $R = \SI{7}{\micro\meter}$. The performance of the pulse $\Fsa(t)$, optimized for non-interacting atoms, quickly deteriorates with increasing interaction strength. Pulses $\Fad (t)$ that are adapted to the interaction fields $\mathbf{I}(t)$ recover most of the performance and deteriorate only for strong interactions. As expected, pulses adapted only to $z$-components $I^z(t)$, i.e., to the frequency shifts induced by interactions, perform comparable if the $\hat{z}\hat{z}$ term in the interactions is dominant (e.g. for $\theta = 0$) and worse if it is negligible, as for $\theta = \text{acos}\left(1/\sqrt{3} \right)$.
   }
 \label{fig: pulse_adaptation}
\end{figure*}

Following Eq.~\eqref{Eq: full effective Hamiltonian}, the interaction fields $\mathbf{I}(t)$ couple to the atoms in the same manner as the pulse $\mathbf{F}(t)$. We can thus interpret the interactions as a disturbance of the pulse, reducing its performance in the circularization. Thereby, the components $I^x(t)$ and $I^y(t)$ generate an amplitude offset, while $I^z(t)$ introduces a detuning. For the atom pair with $R = \SI{7}{\micro\meter}$ and $\theta = \SI{0.1}{\pi}$, the circular state probability reached by $\Fsa (t)$ is reduced from $\pC = \SI{99.0}{\percent}$ for non-interacting atoms to $\mpC = \SI{65.9}{\percent}$, which compares to $\mpC = \SI{65.6}{\percent}$ obtained with the exact model. Our effective model provides a framework to recover high circular state probabilities even with interactions. 

We represent pulses
\begin{align}
\label{Eq: pulses as complex function}
\mathbf{F}(t) = \mathbf{e}_x \Re f(t) + \mathbf{e}_y \Im f(t),
\end{align}
in terms of a complex-valued function $f(t)$ whose real and imaginary parts correspond to $F^x(t)$ and $F^y(t)$. At each time step $k$, we adapt $f_{\text{sa}}(t_k)$ to the interactions $\mathbf{I}(t_k)$ 
\begin{align}
\label{Eq: adapted pulse}
f_{\text{ad}}(t_k) =e^{-i \phi_k}f_{\text{sa}}(t_k) - I^{x}(t_k) - i I^{y}(t_k),
\end{align}
by adding corresponding compensation terms and a time-dependent phase modulation
\begin{align}
\label{Eq: adapted pulse phase}
\phi_k = \Delta t  \frac{3n}{2} \frac{e a_0}{\hbar}\sum_{k' = 1}^k I^{z}(t_{k'}).
\end{align}
We then use the adapted pulse $\Fad(t_k)$ to propagate $\ket{\psi(t_k)}$ to the next time step $k+1$, compute $\mathbf{I}(t_{k+1})$ and repeat the procedure iteratively. This approach allows us to obtain both the time evolution and a pulse adapted to the interactions through the computation of a single time evolution.

For the atom pair with $R = \SI{7}{\micro\meter}$ and $\theta = \SI{0.1}{\pi}$, the adapted pulse $\Fad (t)$ shown in Fig.~\ref{fig: effective_model} (b) yields a mean circular state probability $\mpC = \SI{99.0}{\percent}$ in the effective model, thus recovering the optimal performance of $\Fsa (t)$. In an exact simulation, $\Fad(t)$ yields a mean circular state probability of $\mpC = \SI{98.3}{\percent}$. The reduction of $\mpC$ increases with the imprecision of the effective model, which in turn grows with the interaction strength. This correlation is illustrated in Fig.~\ref{fig: pulse_adaptation} (a), where we compare the performance of the pulses $\Fsa (t)$ and $\Fad(t)$ for pairs arranged at $\theta = 0$ and interatomic distances $R$ ranging from $\SI{4}{\micro\meter}$ to $\SI{12}{\micro\meter}$.

In an exact simulation, the performance of the pulse $\Fsa (t)$ decreases with smaller interatomic distances falling to $\mpC \leq \SI{1.0}{\percent}$ for $R \leq \SI{5}{\micro\meter}$. The adapted pulses reach $\mpC \geq \SI{95.0}{\percent}$ for moderate interactions $R \geq \SI{5.6}{\micro\meter}$ and any $\theta$. Within the effective model, the adapted pulses completely recover a performance of $\mpC = \SI{99.0}{\percent}$ in the regime $R \geq \SI{7.3}{\micro\meter}$ where the deviation in $\mpC$ with respect to the exact evolution remains below $\SI{0.7}{\percent}$. For smaller $R$ a complete recovery is no longer possible, reaching $\mpC = \SI{96.2}{\percent}$ for $R = \SI{4}{\micro\meter}$ and $\theta = 0$. This reduction arises from the time-dependent detuning $I^z(t)$ introduced by the interactions. More specifically, the term $\propto \langle \hat{d}^z_j\rangle$ of Eq.~\eqref{eq:interaction_field_z} introduces an anharmonicity in the state space traversed during the circularization that grows with the interaction strength. We compensate for the latter through a time-dependent phase modulation, which can be understood as pulse chirping \cite{Barth2011}. This approach is successful if the anharmonicity is small compared to the strength of the drive. For strong interactions and correspondingly a large anharmonicity, the frequency of the adapted pulse changes faster than a Rabi cycle of the driven transitions, such that the system cannot remain phase-locked to the drive and coherent population transfer breaks down. Therefore, the phase modulation that compensates the detuning $I^z(t)$ fundamentally limits the pulse adaptation method.

If no phase modulation is required, adapted pulses can recover $\mpC = \SI{99.0}{\percent}$ in the effective model even for strong interactions at $R = \SI{4}{\micro\meter}$. This case is shown in Fig.~\ref{fig: pulse_adaptation} (b), where we compare pairs arranged at the angle $\theta = \acos(1/\sqrt{3})$ for which the leading term of the interactions is canceled and the anharmonicity introduced by $I^z(t)$ vanishes. In this case, the pulse adaptation introduces only amplitude compensations and is consequently only limited by amplitude constraints. The pulse $\Fsa(t)$ is optimized for single atoms such that its amplitudes $F_{\text{sa}}^{x(y)}(t)$ remain below $\SI{92}{\milli\volt\per\centi\meter}$ and its frequencies below $\SI{480}{\mega\hertz}$ (for details on the implementation see Ref. \cite{Huels26}). The adapted pulse $\Fad (t)$, however, may surpass the latter bounds due to the compensations induced by the interaction fields $\mathbf{I}(t)$. While the frequency compensations remain small (yielding shifts of at most $\SI{10}{\mega\hertz}$ for interatomic distances above $\SI{4}{\micro\meter}$) the amplitude compensations can be significant, reaching up to $\SI{72}{\milli\volt\per\centi\meter}$ for an atom pair arranged at $R = \SI{4}{\micro\meter}$ and $\theta = \SI{0.25}{\pi}$. Note that we have chosen a coordinate frame where the interatomic axis lies in the $x-z$ plane, such that significant amplitude compensations are applied to the $x$-component of the pulses $\Fad (t)$ only, see Fig.~\ref{fig: effective_model} (a). For an experimental implementation the interatomic axis should ideally align along $(\mathbf{e}_x+\mathbf{e}_y)/\sqrt{2}$, in which case a pair arranged at $R = \SI{4}{\micro\meter}$ and $\theta = \SI{0.25}{\pi}$ requires amplitude compensations of at most $\SI{57}{\milli\volt\per\centi\meter}$ in both $F^x_{\text{ad}}(t)$ and $F^y_{\text{ad}}(t)$. This equal distribution over both components reduces the performance loss induced by enforcing amplitude constraints.

Alternatively, we could neglect amplitude compensations and adapt to detunings $I^z(t)$ only \cite{Huels26}. The resulting pulses $\Fzad (t)$ always satisfy the amplitude constraints and perform comparably to the fully adapted pulses $\Fad(t)$, as shown in Fig.~\ref{fig: pulse_adaptation}. The difference in the mean circular state probabilities reached is below $\SI{0.5}{\percent}$ for atom pairs separated by $R = \SI{7}{\micro\meter}$. This difference maximizes around the angle $\theta = \thetam$, where the detuning introduced by $I^z(t)$ vanishes. The latter indicates that the detunings $I^z(t)$, which bring the single particle pulse $\Fsa(t)$ out of resonance, constitute the dominant effect by which the interactions disturb the circularization. The pulses $\Fzad (t)$ perform similarly to the pulses obtained using the optimization scheme of Ref.~\cite{Huels26}, where $\Fsa(t)$ is adapted to frequency shifts by optimizing at most 2 parameters. In contrast, the pulses $\Fad (t)$ include amplitude compensations and are thus more general and performative. Adapted pulses can be further improved using pulse-shaping algorithms \cite{Patsch2018, Huels26}. 

Overall, adapting the pulse $\Fsa (t)$ to moderate interactions of an atom pair largely recovers its initial performance of $\mpC = \SI{99.0}{\percent}$. We can follow a similar strategy for any arrangement of multiple atoms. However, in contrast to an atom pair, a system of $N$ atoms is in general not symmetric under a permutation of atoms. This asymmetry is reflected in the circular state probabilities $p_{\text{C},i}$ and effective interaction fields $\mathbf{I}_i(t)$ that differ for different atoms $i$, see Eq.~\eqref{Eq: N Atom effective interaction Hamiltonian}. The RF pulse, however, is a global control that couples to each atom similarly, preventing a local adaptation to $\mathbf{I}_i(t)$. The asymmetry of the interactions thus prevents a simultaneous resonance of the pulses with all atoms. As a compromise, we could maximize the mean circular state probability of all atoms $\bar{p}_C$ by adapting the pulse $\Fsa(t)$ to the mean interaction fields
\begin{align}
\label{Eq: Mean interactions field}
\bar{\mathbf{I}}(t) = \frac{1}{N}\sum_{i=1}^N \mathbf{I}_{i}(t). 
\end{align}
This approach will lead to high (low) circular state probabilities for atoms whose interaction fields $\mathbf{I}_i(t)$ lie close to (far from) the mean $\bar{\mathbf{I}}(t)$. The mean performance $\bar{p}_C$ of the adapted pulse thus strongly depends on the degree of symmetry in the system. Alternatively, we can focus on an atom $i$ and prioritize a maximization of $p_{\text{C},i}$ by adapting $\Fsa (t)$ to $\mathbf{I}_i(t)$. The focus may also lie on a subset of atoms for which the symmetry is large, i.e. the fields $\mathbf{I}_i(t)$ do not differ much. In large arrays of atoms, this subset could correspond, for example, to the bulk of the array or a (computational) sub-lattice. All in all, the asymmetry remains a fundamental problem for the circularization. It may be overcome by arranging atoms in a way that maximizes the symmetry of the system. A square arrangement of four atoms, for example, can be arranged such that a complete permutation symmetry is restored. For other arrangements, one could optimize remaining degrees of freedom, such as the orientation of the static fields $\mathbfcal{E}$ and $\mathbf{B}$ to maximize the symmetry. We are currently working on these directions.

In conclusion, we propose an effective, semi-classical model to simulate the circularization of arbitrary arrays of $N$ interacting atoms. The model relies on a mean field approximation, assuming that each atom evolves in a classical field generated by the interactions with remaining atoms. A self-consistent time evolution is obtained by propagating all atoms in parallel within a state space dimension $d$ per atom, leading to an effective total state space dimension of $Nd$. For pairs of atoms prepared in $n=52$ Rydberg states and separated by $R \geq \SI{7}{\micro\meter}$, the evolved states obtained from the effective model overlap by at least $\SI{99}{\percent}$ with those from the exact model. Deviations from the exact simulation are due to the entanglement among atoms that is neglected within the effective model. Besides an efficient simulation, the effective model allows us to adapt a pulse $\Fsa(t)$ optimized for non-interacting atoms to the interactions. By adding suitable compensations in terms of amplitude and phase modulations, the resulting adapted pulse $\Fad (t)$ recovers the initial performance of $\Fsa(t)$ in the effective model. Adapted pulses can be computed \textit{on the fly} in the course of a single time evolution. By applying compensation terms that exactly cancel the effect of interactions within a mean-field approximation, our adapted pulses achieve a remarkable performance in the regime of weak to moderate interaction strengths where the entanglement among atoms can be neglected. For atom pairs separated by $R \geq \SI{7}{\micro\meter}$, they lead to mean circular pair state probabilities $\mpC \geq \SI{98}{\percent}$ in an exact simulation. In the regime of strong interactions, the adapted pulses can serve as an initial guess for a subsequent optimization performed by pulse-shaping algorithms with access to an exact simulation or a closed feedback loop with the experiment.

\begin{acknowledgments}
We acknowledge funding from the Horizon Europe program HORIZON-CL4-2022-QUANTUM-02-SGA via the project \href{https://doi.org/10.3030/101113690}{101113690} (PASQuanS2.1) and Germany’s Excellence Strategy through the Cluster of Excellence Matter and Light for Quantum Computing
(ML4Q2), EXC 2004/2 – 390534769. E. Cuestas was supported by JSPS KAKENHI grant number JP23K13035. We are grateful to Michel Brune, Clément Sayrin, Aurore A. Young and Jean-Michel Raimond for stimulating discussions that have been instrumental in deepening our understanding of circular Rydberg atoms. We further thank Robert Zeier for useful comments. During the preparation of this work, the authors used the generative AI tools GitHub Copilot and ChatGPT for code debugging, code optimization, and to improve the readability of written text. The authors validated all generated content and take full responsibility for the content of the manuscript.
\end{acknowledgments}

\bibliography{ry_circ_bib}

\begin{thebibliography}{20}%
\makeatletter
\providecommand \@ifxundefined [1]{%
 \@ifx{#1\undefined}
}%
\providecommand \@ifnum [1]{%
 \ifnum #1\expandafter \@firstoftwo
 \else \expandafter \@secondoftwo
 \fi
}%
\providecommand \@ifx [1]{%
 \ifx #1\expandafter \@firstoftwo
 \else \expandafter \@secondoftwo
 \fi
}%
\providecommand \natexlab [1]{#1}%
\providecommand \enquote  [1]{``#1''}%
\providecommand \bibnamefont  [1]{#1}%
\providecommand \bibfnamefont [1]{#1}%
\providecommand \citenamefont [1]{#1}%
\providecommand \href@noop [0]{\@secondoftwo}%
\providecommand \href [0]{\begingroup \@sanitize@url \@href}%
\providecommand \@href[1]{\@@startlink{#1}\@@href}%
\providecommand \@@href[1]{\endgroup#1\@@endlink}%
\providecommand \@sanitize@url [0]{\catcode `\\12\catcode `\$12\catcode
  `\&12\catcode `\#12\catcode `\^12\catcode `\_12\catcode `\%12\relax}%
\providecommand \@@startlink[1]{}%
\providecommand \@@endlink[0]{}%
\providecommand \url  [0]{\begingroup\@sanitize@url \@url }%
\providecommand \@url [1]{\endgroup\@href {#1}{\urlprefix }}%
\providecommand \urlprefix  [0]{URL }%
\providecommand \Eprint [0]{\href }%
\providecommand \doibase [0]{https://doi.org/}%
\providecommand \selectlanguage [0]{\@gobble}%
\providecommand \bibinfo  [0]{\@secondoftwo}%
\providecommand \bibfield  [0]{\@secondoftwo}%
\providecommand \translation [1]{[#1]}%
\providecommand \BibitemOpen [0]{}%
\providecommand \bibitemStop [0]{}%
\providecommand \bibitemNoStop [0]{.\EOS\space}%
\providecommand \EOS [0]{\spacefactor3000\relax}%
\providecommand \BibitemShut  [1]{\csname bibitem#1\endcsname}%
\let\auto@bib@innerbib\@empty
\bibitem [{\citenamefont {H\"{o}lzl}\ \emph {et~al.}(2024)\citenamefont
  {H\"{o}lzl}, \citenamefont {G\"{o}tzelmann}, \citenamefont {Pultinevicius},
  \citenamefont {Wirth},\ and\ \citenamefont {Meinert}}]{Hlzl2024}%
  \BibitemOpen
  \bibfield  {author} {\bibinfo {author} {\bibfnamefont {C.}~\bibnamefont
  {H\"{o}lzl}}, \bibinfo {author} {\bibfnamefont {A.}~\bibnamefont
  {G\"{o}tzelmann}}, \bibinfo {author} {\bibfnamefont {E.}~\bibnamefont
  {Pultinevicius}}, \bibinfo {author} {\bibfnamefont {M.}~\bibnamefont
  {Wirth}},\ and\ \bibinfo {author} {\bibfnamefont {F.}~\bibnamefont
  {Meinert}},\ }\bibfield  {journal} {\bibinfo  {journal} {Physical Review X}\
  }\textbf {\bibinfo {volume} {14}},\ \href
  {https://doi.org/10.1103/physrevx.14.021024} {10.1103/physrevx.14.021024}
  (\bibinfo {year} {2024})\BibitemShut {NoStop}%
\bibitem [{\citenamefont {Wu}\ \emph {et~al.}(2023)\citenamefont {Wu},
  \citenamefont {Richaud}, \citenamefont {Raimond}, \citenamefont {Brune},\
  and\ \citenamefont {Gleyzes}}]{Wu2023}%
  \BibitemOpen
  \bibfield  {author} {\bibinfo {author} {\bibfnamefont {H.}~\bibnamefont
  {Wu}}, \bibinfo {author} {\bibfnamefont {R.}~\bibnamefont {Richaud}},
  \bibinfo {author} {\bibfnamefont {J.-M.}\ \bibnamefont {Raimond}}, \bibinfo
  {author} {\bibfnamefont {M.}~\bibnamefont {Brune}},\ and\ \bibinfo {author}
  {\bibfnamefont {S.}~\bibnamefont {Gleyzes}},\ }\bibfield  {journal} {\bibinfo
   {journal} {Physical Review Letters}\ }\textbf {\bibinfo {volume} {130}},\
  \href {https://doi.org/10.1103/physrevlett.130.023202}
  {10.1103/physrevlett.130.023202} (\bibinfo {year} {2023})\BibitemShut
  {NoStop}%
\bibitem [{\citenamefont {Ravon}\ \emph {et~al.}(2023)\citenamefont {Ravon},
  \citenamefont {Méhaignerie}, \citenamefont {Machu}, \citenamefont
  {Hernández}, \citenamefont {Favier}, \citenamefont {Raimond}, \citenamefont
  {Brune},\ and\ \citenamefont {Sayrin}}]{Ravon2023}%
  \BibitemOpen
  \bibfield  {author} {\bibinfo {author} {\bibfnamefont {B.}~\bibnamefont
  {Ravon}}, \bibinfo {author} {\bibfnamefont {P.}~\bibnamefont {Méhaignerie}},
  \bibinfo {author} {\bibfnamefont {Y.}~\bibnamefont {Machu}}, \bibinfo
  {author} {\bibfnamefont {A.~D.}\ \bibnamefont {Hernández}}, \bibinfo
  {author} {\bibfnamefont {M.}~\bibnamefont {Favier}}, \bibinfo {author}
  {\bibfnamefont {J.~M.}\ \bibnamefont {Raimond}}, \bibinfo {author}
  {\bibfnamefont {M.}~\bibnamefont {Brune}},\ and\ \bibinfo {author}
  {\bibfnamefont {C.}~\bibnamefont {Sayrin}},\ }\bibfield  {journal} {\bibinfo
  {journal} {Physical Review Letters}\ }\textbf {\bibinfo {volume} {131}},\
  \href {https://doi.org/10.1103/physrevlett.131.093401}
  {10.1103/physrevlett.131.093401} (\bibinfo {year} {2023})\BibitemShut
  {NoStop}%
\bibitem [{\citenamefont {Cohen}\ and\ \citenamefont
  {Thompson}(2021)}]{Cohen2021}%
  \BibitemOpen
  \bibfield  {author} {\bibinfo {author} {\bibfnamefont {S.~R.}\ \bibnamefont
  {Cohen}}\ and\ \bibinfo {author} {\bibfnamefont {J.~D.}\ \bibnamefont
  {Thompson}},\ }\bibfield  {journal} {\bibinfo  {journal} {PRX Quantum}\
  }\textbf {\bibinfo {volume} {2}},\ \href
  {https://doi.org/10.1103/prxquantum.2.030322} {10.1103/prxquantum.2.030322}
  (\bibinfo {year} {2021})\BibitemShut {NoStop}%
\bibitem [{\citenamefont {Nguyen}\ \emph {et~al.}(2018)\citenamefont {Nguyen},
  \citenamefont {Raimond}, \citenamefont {Sayrin}, \citenamefont {Cortiñas},
  \citenamefont {Cantat-Moltrecht}, \citenamefont {Assemat}, \citenamefont
  {Dotsenko}, \citenamefont {Gleyzes}, \citenamefont {Haroche}, \citenamefont
  {Roux}, \citenamefont {Jolicoeur},\ and\ \citenamefont {Brune}}]{Nguyen2018}%
  \BibitemOpen
  \bibfield  {author} {\bibinfo {author} {\bibfnamefont {T.}~\bibnamefont
  {Nguyen}}, \bibinfo {author} {\bibfnamefont {J.}~\bibnamefont {Raimond}},
  \bibinfo {author} {\bibfnamefont {C.}~\bibnamefont {Sayrin}}, \bibinfo
  {author} {\bibfnamefont {R.}~\bibnamefont {Cortiñas}}, \bibinfo {author}
  {\bibfnamefont {T.}~\bibnamefont {Cantat-Moltrecht}}, \bibinfo {author}
  {\bibfnamefont {F.}~\bibnamefont {Assemat}}, \bibinfo {author} {\bibfnamefont
  {I.}~\bibnamefont {Dotsenko}}, \bibinfo {author} {\bibfnamefont
  {S.}~\bibnamefont {Gleyzes}}, \bibinfo {author} {\bibfnamefont
  {S.}~\bibnamefont {Haroche}}, \bibinfo {author} {\bibfnamefont
  {G.}~\bibnamefont {Roux}}, \bibinfo {author} {\bibfnamefont {T.}~\bibnamefont
  {Jolicoeur}},\ and\ \bibinfo {author} {\bibfnamefont {M.}~\bibnamefont
  {Brune}},\ }\bibfield  {journal} {\bibinfo  {journal} {Physical Review X}\
  }\textbf {\bibinfo {volume} {8}},\ \href
  {https://doi.org/10.1103/physrevx.8.011032} {10.1103/physrevx.8.011032}
  (\bibinfo {year} {2018})\BibitemShut {NoStop}%
\bibitem [{\citenamefont {Meinert}\ \emph {et~al.}(2020)\citenamefont
  {Meinert}, \citenamefont {H\"{o}lzl}, \citenamefont {Nebioglu}, \citenamefont
  {D’Arnese}, \citenamefont {Karl}, \citenamefont {Dressel},\ and\
  \citenamefont {Scheffler}}]{Meinert2020}%
  \BibitemOpen
  \bibfield  {author} {\bibinfo {author} {\bibfnamefont {F.}~\bibnamefont
  {Meinert}}, \bibinfo {author} {\bibfnamefont {C.}~\bibnamefont {H\"{o}lzl}},
  \bibinfo {author} {\bibfnamefont {M.~A.}\ \bibnamefont {Nebioglu}}, \bibinfo
  {author} {\bibfnamefont {A.}~\bibnamefont {D’Arnese}}, \bibinfo {author}
  {\bibfnamefont {P.}~\bibnamefont {Karl}}, \bibinfo {author} {\bibfnamefont
  {M.}~\bibnamefont {Dressel}},\ and\ \bibinfo {author} {\bibfnamefont
  {M.}~\bibnamefont {Scheffler}},\ }\bibfield  {journal} {\bibinfo  {journal}
  {Physical Review Research}\ }\textbf {\bibinfo {volume} {2}},\ \href
  {https://doi.org/10.1103/physrevresearch.2.023192}
  {10.1103/physrevresearch.2.023192} (\bibinfo {year} {2020})\BibitemShut
  {NoStop}%
\bibitem [{\citenamefont {Facon}\ \emph {et~al.}(2016)\citenamefont {Facon},
  \citenamefont {Dietsche}, \citenamefont {Grosso}, \citenamefont {Haroche},
  \citenamefont {Raimond}, \citenamefont {Brune},\ and\ \citenamefont
  {Gleyzes}}]{Facon2016}%
  \BibitemOpen
  \bibfield  {author} {\bibinfo {author} {\bibfnamefont {A.}~\bibnamefont
  {Facon}}, \bibinfo {author} {\bibfnamefont {E.-K.}\ \bibnamefont {Dietsche}},
  \bibinfo {author} {\bibfnamefont {D.}~\bibnamefont {Grosso}}, \bibinfo
  {author} {\bibfnamefont {S.}~\bibnamefont {Haroche}}, \bibinfo {author}
  {\bibfnamefont {J.-M.}\ \bibnamefont {Raimond}}, \bibinfo {author}
  {\bibfnamefont {M.}~\bibnamefont {Brune}},\ and\ \bibinfo {author}
  {\bibfnamefont {S.}~\bibnamefont {Gleyzes}},\ }\href
  {https://doi.org/10.1038/nature18327} {\bibfield  {journal} {\bibinfo
  {journal} {Nature}\ }\textbf {\bibinfo {volume} {535}},\ \bibinfo {pages}
  {262–265} (\bibinfo {year} {2016})}\BibitemShut {NoStop}%
\bibitem [{\citenamefont {Dietsche}\ \emph {et~al.}(2019)\citenamefont
  {Dietsche}, \citenamefont {Larrouy}, \citenamefont {Haroche}, \citenamefont
  {Raimond}, \citenamefont {Brune},\ and\ \citenamefont
  {Gleyzes}}]{Dietsche2019}%
  \BibitemOpen
  \bibfield  {author} {\bibinfo {author} {\bibfnamefont {E.~K.}\ \bibnamefont
  {Dietsche}}, \bibinfo {author} {\bibfnamefont {A.}~\bibnamefont {Larrouy}},
  \bibinfo {author} {\bibfnamefont {S.}~\bibnamefont {Haroche}}, \bibinfo
  {author} {\bibfnamefont {J.~M.}\ \bibnamefont {Raimond}}, \bibinfo {author}
  {\bibfnamefont {M.}~\bibnamefont {Brune}},\ and\ \bibinfo {author}
  {\bibfnamefont {S.}~\bibnamefont {Gleyzes}},\ }\href
  {https://doi.org/10.1038/s41567-018-0405-4} {\bibfield  {journal} {\bibinfo
  {journal} {Nature Physics}\ }\textbf {\bibinfo {volume} {15}},\ \bibinfo
  {pages} {326–329} (\bibinfo {year} {2019})}\BibitemShut {NoStop}%
\bibitem [{\citenamefont {Nussenzveig}\ \emph {et~al.}(1993)\citenamefont
  {Nussenzveig}, \citenamefont {Bernardot}, \citenamefont {Brune},
  \citenamefont {Hare}, \citenamefont {Raimond}, \citenamefont {Haroche},\ and\
  \citenamefont {Gawlik}}]{Nussenzveig1993}%
  \BibitemOpen
  \bibfield  {author} {\bibinfo {author} {\bibfnamefont {P.}~\bibnamefont
  {Nussenzveig}}, \bibinfo {author} {\bibfnamefont {F.}~\bibnamefont
  {Bernardot}}, \bibinfo {author} {\bibfnamefont {M.}~\bibnamefont {Brune}},
  \bibinfo {author} {\bibfnamefont {J.}~\bibnamefont {Hare}}, \bibinfo {author}
  {\bibfnamefont {J.~M.}\ \bibnamefont {Raimond}}, \bibinfo {author}
  {\bibfnamefont {S.}~\bibnamefont {Haroche}},\ and\ \bibinfo {author}
  {\bibfnamefont {W.}~\bibnamefont {Gawlik}},\ }\href
  {https://doi.org/10.1103/physreva.48.3991} {\bibfield  {journal} {\bibinfo
  {journal} {Physical Review A}\ }\textbf {\bibinfo {volume} {48}},\ \bibinfo
  {pages} {3991–3994} (\bibinfo {year} {1993})}\BibitemShut {NoStop}%
\bibitem [{\citenamefont {Signoles}\ \emph {et~al.}(2017)\citenamefont
  {Signoles}, \citenamefont {Dietsche}, \citenamefont {Facon}, \citenamefont
  {Grosso}, \citenamefont {Haroche}, \citenamefont {Raimond}, \citenamefont
  {Brune},\ and\ \citenamefont {Gleyzes}}]{Signoles2017}%
  \BibitemOpen
  \bibfield  {author} {\bibinfo {author} {\bibfnamefont {A.}~\bibnamefont
  {Signoles}}, \bibinfo {author} {\bibfnamefont {E.}~\bibnamefont {Dietsche}},
  \bibinfo {author} {\bibfnamefont {A.}~\bibnamefont {Facon}}, \bibinfo
  {author} {\bibfnamefont {D.}~\bibnamefont {Grosso}}, \bibinfo {author}
  {\bibfnamefont {S.}~\bibnamefont {Haroche}}, \bibinfo {author} {\bibfnamefont
  {J.}~\bibnamefont {Raimond}}, \bibinfo {author} {\bibfnamefont
  {M.}~\bibnamefont {Brune}},\ and\ \bibinfo {author} {\bibfnamefont
  {S.}~\bibnamefont {Gleyzes}},\ }\bibfield  {journal} {\bibinfo  {journal}
  {Physical Review Letters}\ }\textbf {\bibinfo {volume} {118}},\ \href
  {https://doi.org/10.1103/physrevlett.118.253603}
  {10.1103/physrevlett.118.253603} (\bibinfo {year} {2017})\BibitemShut
  {NoStop}%
\bibitem [{\citenamefont {Patsch}\ \emph {et~al.}(2018)\citenamefont {Patsch},
  \citenamefont {Reich}, \citenamefont {Raimond}, \citenamefont {Brune},
  \citenamefont {Gleyzes},\ and\ \citenamefont {Koch}}]{Patsch2018}%
  \BibitemOpen
  \bibfield  {author} {\bibinfo {author} {\bibfnamefont {S.}~\bibnamefont
  {Patsch}}, \bibinfo {author} {\bibfnamefont {D.~M.}\ \bibnamefont {Reich}},
  \bibinfo {author} {\bibfnamefont {J.-M.}\ \bibnamefont {Raimond}}, \bibinfo
  {author} {\bibfnamefont {M.}~\bibnamefont {Brune}}, \bibinfo {author}
  {\bibfnamefont {S.}~\bibnamefont {Gleyzes}},\ and\ \bibinfo {author}
  {\bibfnamefont {C.~P.}\ \bibnamefont {Koch}},\ }\bibfield  {journal}
  {\bibinfo  {journal} {Physical Review A}\ }\textbf {\bibinfo {volume} {97}},\
  \href {https://doi.org/10.1103/physreva.97.053418}
  {10.1103/physreva.97.053418} (\bibinfo {year} {2018})\BibitemShut {NoStop}%
\bibitem [{\citenamefont {Larrouy}\ \emph {et~al.}(2020)\citenamefont
  {Larrouy}, \citenamefont {Patsch}, \citenamefont {Richaud}, \citenamefont
  {Raimond}, \citenamefont {Brune}, \citenamefont {Koch},\ and\ \citenamefont
  {Gleyzes}}]{Larrouy2020}%
  \BibitemOpen
  \bibfield  {author} {\bibinfo {author} {\bibfnamefont {A.}~\bibnamefont
  {Larrouy}}, \bibinfo {author} {\bibfnamefont {S.}~\bibnamefont {Patsch}},
  \bibinfo {author} {\bibfnamefont {R.}~\bibnamefont {Richaud}}, \bibinfo
  {author} {\bibfnamefont {J.-M.}\ \bibnamefont {Raimond}}, \bibinfo {author}
  {\bibfnamefont {M.}~\bibnamefont {Brune}}, \bibinfo {author} {\bibfnamefont
  {C.~P.}\ \bibnamefont {Koch}},\ and\ \bibinfo {author} {\bibfnamefont
  {S.}~\bibnamefont {Gleyzes}},\ }\bibfield  {journal} {\bibinfo  {journal}
  {Physical Review X}\ }\textbf {\bibinfo {volume} {10}},\ \href
  {https://doi.org/10.1103/physrevx.10.021058} {10.1103/physrevx.10.021058}
  (\bibinfo {year} {2020})\BibitemShut {NoStop}%
\bibitem [{\citenamefont {Méhaignerie}(2023)}]{mehai2023}%
  \BibitemOpen
  \bibfield  {author} {\bibinfo {author} {\bibfnamefont {P.}~\bibnamefont
  {Méhaignerie}},\ }\emph {\bibinfo {title} {Interactions entre atomes de
  {Rydberg} circulaires piégés pour la simulation quantique}},\ \href
  {http://www.theses.fr/2023SORUS259} {Ph.D. thesis},\ \bibinfo  {school}
  {Sorbonne université} (\bibinfo {year} {2023})\BibitemShut {NoStop}%
\bibitem [{\citenamefont {Méhaignerie}\ \emph {et~al.}(2025)\citenamefont
  {Méhaignerie}, \citenamefont {Machu}, \citenamefont {Durán~Hernández},
  \citenamefont {Creutzer}, \citenamefont {Papoular}, \citenamefont {Raimond},
  \citenamefont {Sayrin},\ and\ \citenamefont {Brune}}]{Mhaignerie2025}%
  \BibitemOpen
  \bibfield  {author} {\bibinfo {author} {\bibfnamefont {P.}~\bibnamefont
  {Méhaignerie}}, \bibinfo {author} {\bibfnamefont {Y.}~\bibnamefont {Machu}},
  \bibinfo {author} {\bibfnamefont {A.}~\bibnamefont {Durán~Hernández}},
  \bibinfo {author} {\bibfnamefont {G.}~\bibnamefont {Creutzer}}, \bibinfo
  {author} {\bibfnamefont {D.}~\bibnamefont {Papoular}}, \bibinfo {author}
  {\bibfnamefont {J.}~\bibnamefont {Raimond}}, \bibinfo {author} {\bibfnamefont
  {C.}~\bibnamefont {Sayrin}},\ and\ \bibinfo {author} {\bibfnamefont
  {M.}~\bibnamefont {Brune}},\ }\bibfield  {journal} {\bibinfo  {journal} {PRX
  Quantum}\ }\textbf {\bibinfo {volume} {6}},\ \href
  {https://doi.org/10.1103/prxquantum.6.010353} {10.1103/prxquantum.6.010353}
  (\bibinfo {year} {2025})\BibitemShut {NoStop}%
\bibitem [{\citenamefont {H\"{u}ls}\ \emph {et~al.}(2026)\citenamefont
  {H\"{u}ls}, \citenamefont {Young}, \citenamefont {Sayrin}, \citenamefont
  {Brune}, \citenamefont {Raimond}, \citenamefont {Calarco}, \citenamefont
  {Motzoi}, \citenamefont {Zeier},\ and\ \citenamefont {Cuestas}}]{Huels26}%
  \BibitemOpen
  \bibfield  {author} {\bibinfo {author} {\bibfnamefont {M.}~\bibnamefont
  {H\"{u}ls}}, \bibinfo {author} {\bibfnamefont {A.~A.}\ \bibnamefont {Young}},
  \bibinfo {author} {\bibfnamefont {C.}~\bibnamefont {Sayrin}}, \bibinfo
  {author} {\bibfnamefont {M.}~\bibnamefont {Brune}}, \bibinfo {author}
  {\bibfnamefont {J.-M.}\ \bibnamefont {Raimond}}, \bibinfo {author}
  {\bibfnamefont {T.}~\bibnamefont {Calarco}}, \bibinfo {author} {\bibfnamefont
  {F.}~\bibnamefont {Motzoi}}, \bibinfo {author} {\bibfnamefont
  {R.}~\bibnamefont {Zeier}},\ and\ \bibinfo {author} {\bibfnamefont
  {E.}~\bibnamefont {Cuestas}},\ }\href
  {https://doi.org/10.48550/ARXIV.2607.05216} {\bibinfo {title} {Fast pulses
  for high-fidelity circularization of interacting {Rydberg} atoms}} (\bibinfo
  {year} {2026})\BibitemShut {NoStop}%
\bibitem [{\citenamefont {Gallagher}(1994)}]{gallagher_1994_book}%
  \BibitemOpen
  \bibfield  {author} {\bibinfo {author} {\bibfnamefont {T.~F.}\ \bibnamefont
  {Gallagher}},\ }\href@noop {} {\emph {\bibinfo {title} {{Rydberg Atoms}}}},\
  Cambridge Monographs on Atomic, Molecular and Chemical Physics\ (\bibinfo
  {publisher} {Cambridge University Press},\ \bibinfo {year}
  {1994})\BibitemShut {NoStop}%
\bibitem [{\citenamefont {Weber}\ \emph {et~al.}(2017)\citenamefont {Weber},
  \citenamefont {Tresp}, \citenamefont {Menke}, \citenamefont {Urvoy},
  \citenamefont {Firstenberg}, \citenamefont {B\"{u}chler},\ and\ \citenamefont
  {Hofferberth}}]{Weber2017}%
  \BibitemOpen
  \bibfield  {author} {\bibinfo {author} {\bibfnamefont {S.}~\bibnamefont
  {Weber}}, \bibinfo {author} {\bibfnamefont {C.}~\bibnamefont {Tresp}},
  \bibinfo {author} {\bibfnamefont {H.}~\bibnamefont {Menke}}, \bibinfo
  {author} {\bibfnamefont {A.}~\bibnamefont {Urvoy}}, \bibinfo {author}
  {\bibfnamefont {O.}~\bibnamefont {Firstenberg}}, \bibinfo {author}
  {\bibfnamefont {H.~P.}\ \bibnamefont {B\"{u}chler}},\ and\ \bibinfo {author}
  {\bibfnamefont {S.}~\bibnamefont {Hofferberth}},\ }\href
  {https://doi.org/10.1088/1361-6455/aa743a} {\bibfield  {journal} {\bibinfo
  {journal} {Journal of Physics B: Atomic, Molecular and Optical Physics}\
  }\textbf {\bibinfo {volume} {50}},\ \bibinfo {pages} {133001} (\bibinfo
  {year} {2017})}\BibitemShut {NoStop}%
\bibitem [{\citenamefont {Ravets}\ \emph {et~al.}(2015)\citenamefont {Ravets},
  \citenamefont {Labuhn}, \citenamefont {Barredo}, \citenamefont {Lahaye},\
  and\ \citenamefont {Browaeys}}]{Ravets2015}%
  \BibitemOpen
  \bibfield  {author} {\bibinfo {author} {\bibfnamefont {S.}~\bibnamefont
  {Ravets}}, \bibinfo {author} {\bibfnamefont {H.}~\bibnamefont {Labuhn}},
  \bibinfo {author} {\bibfnamefont {D.}~\bibnamefont {Barredo}}, \bibinfo
  {author} {\bibfnamefont {T.}~\bibnamefont {Lahaye}},\ and\ \bibinfo {author}
  {\bibfnamefont {A.}~\bibnamefont {Browaeys}},\ }\bibfield  {journal}
  {\bibinfo  {journal} {Physical Review A}\ }\textbf {\bibinfo {volume} {92}},\
  \href {https://doi.org/10.1103/physreva.92.020701}
  {10.1103/physreva.92.020701} (\bibinfo {year} {2015})\BibitemShut {NoStop}%
\bibitem [{\citenamefont {Hwang}\ \emph {et~al.}(2023)\citenamefont {Hwang},
  \citenamefont {Byun}, \citenamefont {Park}, \citenamefont
  {de~L\'{e}s\'{e}leuc},\ and\ \citenamefont {Ahn}}]{Hwang23}%
  \BibitemOpen
  \bibfield  {author} {\bibinfo {author} {\bibfnamefont {H.}~\bibnamefont
  {Hwang}}, \bibinfo {author} {\bibfnamefont {A.}~\bibnamefont {Byun}},
  \bibinfo {author} {\bibfnamefont {J.}~\bibnamefont {Park}}, \bibinfo {author}
  {\bibfnamefont {S.}~\bibnamefont {de~L\'{e}s\'{e}leuc}},\ and\ \bibinfo
  {author} {\bibfnamefont {J.}~\bibnamefont {Ahn}},\ }\href
  {https://doi.org/10.1364/OPTICA.480535} {\bibfield  {journal} {\bibinfo
  {journal} {Optica}\ }\textbf {\bibinfo {volume} {10}},\ \bibinfo {pages}
  {401} (\bibinfo {year} {2023})}\BibitemShut {NoStop}%
\bibitem [{\citenamefont {Barth}\ \emph {et~al.}(2011)\citenamefont {Barth},
  \citenamefont {Friedland}, \citenamefont {Gat},\ and\ \citenamefont
  {Shagalov}}]{Barth2011}%
  \BibitemOpen
  \bibfield  {author} {\bibinfo {author} {\bibfnamefont {I.}~\bibnamefont
  {Barth}}, \bibinfo {author} {\bibfnamefont {L.}~\bibnamefont {Friedland}},
  \bibinfo {author} {\bibfnamefont {O.}~\bibnamefont {Gat}},\ and\ \bibinfo
  {author} {\bibfnamefont {A.~G.}\ \bibnamefont {Shagalov}},\ }\bibfield
  {journal} {\bibinfo  {journal} {Physical Review A}\ }\textbf {\bibinfo
  {volume} {84}},\ \href {https://doi.org/10.1103/physreva.84.013837}
  {10.1103/physreva.84.013837} (\bibinfo {year} {2011})\BibitemShut {NoStop}%
\end{thebibliography}%
\bibliographystyle{apsrev4-2}
\end{document}